\documentclass[a4paper]{article}
\usepackage{graphicx,amsmath,amssymb}

\newcommand{\rhaak}[1]{\!\left [#1\right]}
\newcommand{\lhaak}[1]{\left | #1\right |}

\newcommand{\gemr}[1]{\left. #1\right\rangle}

\newcommand{\lhaakl}[1]{\left |#1\right.}

\newcommand{\ket}[1]{\lhaakl{\gemr{#1}}}

\newcommand{\brak}[2]{\left\langle #1\vphantom{#2}\right | \!\left.\vphantom{#1}#2
\right\rangle}

\newcommand{\bfm}[1]{\mathbf{#1}}
\renewcommand{\imath}{\text{i}}

\begin{document}
\title{Entangled states considered as physical representations of classical algorithms} 
\author{Saibal Mitra}
\date{\today}
\maketitle
\begin{abstract}
There are good reasons to believe that we are classical algorithms run on (effectively) classical machines. However, the fact that a physical state of a system in a universe described by a classical deterministic model doesn't contain any information about the model's evolution laws, gives rise to deep philosophical paradoxes with this picture of what we are. We explain these paradoxes in detail and show that they can be resolved once we take into account that in the real world, classical behavior arises as a result of decoherence. We then show that this solution naturally leads to a variant of the idea of a mathematical multiverse, originally proposed by Tegmark.

\end{abstract}

\section{Introduction}
According to the computationalist theory of the mind, conscious experiences are identified with  computational states of algorithms \cite{compt,compb}. This view is the logical conclusion one arrives at if one assumes that physics applies to everything, including us. One can then still question if classical computation can fully explain conscious experiences generated by the brain. Arguments have been put forward that suggest that quantum computing may play an important role in some brain processes \cite{quantbrain}. However, Tegmark has shown that the relevant time scale for elementary information processing events in the brain to unfold, is many orders of magnitude larger than the time it takes for superpositions of such processes to decohere \cite{tegbrain}. This means that whatever human consciousness is, it has to be identified with the running of certain classical algorithms. Any randomness due to quantum mechanics manifests itself in a purely classical way (i.e. without leading to interference phenomena or violations of Bell's inequalities) and can thus always be modeled by pseudo-random generators \cite{tegbrain}.

Nevertheless, computationalism has been vigorously attacked by some philosophers on many different grounds \cite{compb}. Many of the criticisms boil down to computationalism not being compatible with some prior assumptions made about consciousness. In this article we won't consider such arguments. We will assume the computationalist point of view that conscious experience only depends on the computational state of an algorithm, and is completely independent of the physical implementation. However, even within this framework, some of the criticisms remain relevant. A particularly interesting argument against computationalism is based on the existence of mappings of machine states to the states of trivial systems \cite{crit}.

Consider a simulation of a person in a virtual environment such that the whole evolution of the system is deterministic. A simulation over some finite time interval will simply cause the system to evolve through some finite number of predetermined states. This means that this system behaves just like any trivial system that also evolves through at least the same number of different states, say a clock with a number of independent dials that run at different rates. Why then would such a trivial system not generate the same subjective experiences? Clearly the time evolutions of the two systems can be mapped onto each other. Nevertheless, the two systems are not the same, but to see this one has to consider a range of different initial conditions. Only then can one see that the computer is actually performing non-trivial computations, while the clock is just running in the same trivial way \cite{rep}. No one to one mapping between the states of the systems is possible when one considers counterfactual initial conditions. 

While it is a compelling argument that counterfactuals must be considered to see if a computation is really being implemented by a system, one still has to deal with the fact that counterfactuals are events that, by definition, did not happen. To see that this is not just a philosophical problem with no relevance to physics, we can reformulate it into one where the same paradox manifests itself as ambiguities in probabilities of experimental outcomes \cite{alm}. Suppose we run three simultaneous simulations of a person, one in a virtual room with white walls and two simulations in a virtual room with black walls. The latter two simulations are exactly identical; the processors will behave in exactly the same way. The probability that the person finds himself in the black room would seem to be $2/3$. However, since the two processors rendering the person in the black room are performing identical operations, we could imagine replacing one processor by a "dummy processor" that simply copies whatever the other one is doing. Since no real computations are performed by the dummy processor, we should expect that the probabilities are now $1/2$. In terms of counterfactuals, we can say that when both processors are working independently, one can choose independent input for both systems. But then a similar objection can be made as in the previous case: In a situation where both processors are going to perform identical computations, why does it matter if both processors can function correctly when fed different input?

The fundamental problem these paradoxes point to is that in classical deterministic models, the information about the evolution laws of the system is not present in the physical states of the system itself. While this poses no problems when doing computations in terms of the physical states, it becomes a problem when we imagine that an observer is present in a universe described by such deterministic laws and we attempt to compute the probabilities of certain observations. However, we've only seen that this is a problem for a hypothetical world that is exactly described by a classical deterministic model. While the real world at the macro-level is effectively classical in the sense that probabilities behave in a classical way, i.e.\ without exhibiting interference or violations of Bell's inequalities, the reason why the above paradoxes arise has nothing to do with classicality in this sense. It thus makes sense to re-examine the paradoxes in  a real world setting.

\section{Classical algorithms run on real machines}
The world we live in is described by quantum mechanics. This means that even macroscopic machines running classical algorithms are described by quantum mechanics. The classical behavior of such a machine is explained by decoherence: On a time scale that is much smaller than the time needed to perform an elementary computational step, any superposition of machine states would have decohered; in fact such superpositions won't arise due to fast decoherence in the first place. As mentioned in the previous section, we note that Tegmark has shown that this picture is adequate for the human brain \cite{tegbrain}.

The computational state a classical machine is in, can be specified by a bit string $\bfm{b}$. The quantum states corresponding to these bit strings, which we'll denote as $\ket{\bfm{b}}$, form a subset of the pointer basis. Normalized states of the universe containing the machine can be denoted as
\begin{equation}\label{state}
\ket{\psi}=\sum_{\bfm{b}}\ket{\bfm{b}}\otimes\ket{E(\bfm{b})}
\end{equation}
where $\ket{E(\bfm{b})}$ denotes the unnormalized states of the environment which we define as   all the degrees of freedom that are not described by the bit string. We assume that all the environmental states $\ket{E(\bfm{b})}$ describe the machine running the same algorithm reliably; under the time evolution according to the Schr\"{o}dinger equation, the bit string changes after some time step according to the algorithm that the machine is supposed to be running. 

The states $\ket{E(\bfm{b})}$ are orthogonal and thus contain perfect information about the computational state. In fact, the whole computational history of the machine will be contained in the states $\ket{E(\bfm{b})}$. In the Copenhagen and Consistent Histories interpretations, one assumes that only one term of \eqref{state} refers to the real world. Here we won't make any such assumptions, and work within the Many Worlds Interpretation (MWI) \cite{ev}. 

The state \eqref{state} seems to have a straightforward interpretation: According to the Born rule, the probability for the observer generated by the machine to find herself in the computational state $\bfm{b}$ should be $\lhaak{\brak{E(\bfm{b})}{E(\bfm{b})}}^{2}$. However, we would then assume that an observer with some definite experience is described by a single term of \eqref{state}. The only assumption about consciousness we'll make in this article is that the opposite is true: A definite conscious experience is consistent with an astronomically large number of computational states. We can be confident that this must be the case for at least us. A huge amount of brain processes are going on at any given moment. Many of these processes implement pattern recognition algorithms which leads to awareness of the patterns at the expense of awareness of the many parts that make up the patterns \cite{cent}. Since the patterns will have a significant entropy in terms of the computational states, this leads to the conclusion that any particular state of consciousness should be consistent with a large number of computational states.

We can now see how this solves the paradox on counterfactuals discussed in the previous section. An observer with a definite experience is represented by an entangled state of the form \eqref{state}, where the summation is over many computational states. Since the environmental states contain the complete information about the computational history, we can read such an entangled state as a table that specifies the output of the algorithm as a function of the input. The state of the observer thus (partially) specifies the algorithm that generates the observer. Note that interpreting the entangled state in this way is not an ad hoc choice. Compare this to writing down a table on paper specifying each output corresponding to each input. This does not single out the correct interpretation of whatever is written as the algorithm. We may write in the caption how to interpret the table, but whatever is written there is not a law of physics and can thus be ignored.

In contrast, the entangled state \eqref{state} automatically implies that the components of the state are correlated. This correlation does not define the algorithm fully, as the summation over the computational states in the entangled state is only over a restricted set of states that are consistent with whatever the observer is aware of. It is then natural to assume that "awareness" {\it is} such a correlation. Quantum mechanics allows for such correlations to exist in the form of entangled states at any particular moment, while within classical mechanics there is no room for this.

One obvious objection against this picture can be raised: Since the computational states are macroscopic, one can imagine a machine looking at itself and simply observing its own parts. So, why won't the machine be able to observe its own computational state? To answer this, we note that the information the machine has about the external world makes up part of the computational state, therefore the machine cannot have enough memory to store the exact computational state it is in. Moreover, as explained above, it is reasonable to assume that whatever we are aware of, are patterns in the raw information present in the brain, so the number of different states of conscious experiences should be far less than the number of states the memory can encode. We can still imagine a "super-observer" with a memory capacity that is vastly larger than that of the machine that can observe the precise computational state of the machine, but there is then no way that the observation can be communicated to the machine.

\section{A mathematical multiverse?}
The main conclusion we've reached is based on a rather trivial observation. We can take any entangled state, like the following two-qubit state:
\begin{equation}
\frac{1}{\sqrt{2}}\rhaak{\ket{0}\otimes\ket{0} + \ket{1}\otimes\ket{1}}
\end{equation}
and then say that this represents a classical algorithm or a mathematical statement. In this case, we can read the state as saying that the second qubit is in the state $\ket{x}$, if and only if the first qubit is in the state $\ket{x}$. Mathematical statements can thus be said to have objective existence when they are realized in the form of such entangled states. In particular, you are that particular algorithm that your brain is running, and this exists in the form of an entangled state.

The Hamiltonian that determines the dynamics plays a redundant role in this picture. Since we are always localized in time at some given moment, the information we have about the dynamics must be contained in the physical state of the universe at any given time. This suggests that the dynamics may not play the fundamental role in physics as one commonly assumes. The conventional picture may actually not work; it has been hard to find a solution to the so-called Boltzmann brain problem: Contrary to our observations, typical observers in the universe are predicted to arise randomly due to fluctuations and will thus have awareness of false information. But if we simply ignore this problem and focus on a sector containing a normal observer and apply the time evolution operator to predict the probabilities of outcomes of experiments, then no problems occur.

It thus makes sense to throw away the redundancy to prevent it from making trouble. We can do this by postulating the existence of a mathematical multiverse, similar to a proposal by Tegmark \cite{tegmath}. The difference is that what we postulate is simply a set of all possible mathematical statements defining functions or algorithms. Unlike Tegmark, we don't take each element of the ensemble to be a model of a universe in the conventional sense, i.e.\ one which evolves according to some laws. Such laws contain redundant information that we don't want. The elements of the ensemble can be formally represented by quantum states, as explained above. For a given observer in some conscious state, there will be many elements that can represent her. One can e.g.\ append extra qubits to the element, without affecting the interpretation of the algorithm defining the observer. 

One then needs to specify a measure over the ensemble. This fixes a probability distribution over the possible observed laws of physics for every observer in the ensemble. Obviously, the measure must be larger for elements containing information that are compatible with living in a lawful universe. One way to express this mathematically is to say that the information contained in the element must be highly compressible. One can consider the size of the smallest quantum algorithm that can generate the element starting from a reference state (e.g.\ the state in which all qubits initialized to zero, what matters is that this reference state can be specified with a few bits). To get rid of Boltzmann brain observers, one also needs to take into account the number of steps the algorithm needs to be applied. So, what seems to matter is the overall computational complexity of the unitary map from the reference state to the target state.

\section{Discussion}
We have argued that the very mechanism that explains classical behavior of macroscopic systems, i.e.\ decoherence, also solves the paradox of counterfactuals in computationalism. The fact that one can describe macroscopic phenomena using classical mechanics, doesn't make these phenomena classical in the sense that all quantum correlations have magically disappeared. In reality, the macroscopic degrees of freedom get entangled with an astronomically large number of microscopic degrees of freedom, making the macroscopic degrees of freedom appear to behave in a classical way. It is then possible to describe these degrees of freedom using a classical model in which one pretends that the macroscopic degrees of freedom live in a classical phase-space. But this phase-space is unphysical; the real system is still described by a quantum mechanical state vector which in general describes a complicated entangled state. 

We have then argued that this suggests that instead of a physical universe that evolves according to some laws, all that really exists is a mathematical multiverse, whose elements are classical algorithms. These algorithms can be represented formally as quantum states. A measure needs to be defined over this multiverse to get useful physics out of this. We've argued that the measure of an element should depend on the computational complexity of the unitary map that generates the element from a reference state.

\section{Acknowledgments}
I thank Paul Almond and Andrew Soltau for interesting discussions.


\begin{thebibliography}{100}
\bibitem{alm} This thought experiment came up in discussions I had with Paul Almond. Paul raised the paradox, but argued that this shows that the substrate is important in determining probabilities. I made the connection with the paradox on counterfactuals.
\end{thebibliography}
\end{document}